\documentclass{article}

\usepackage[preprint]{neurips_2024}
\usepackage{amsmath}

\usepackage[utf8]{inputenc} % allow utf-8 input
\usepackage[T1]{fontenc}    % use 8-bit T1 fonts
\usepackage{hyperref}       % hyperlinks
\usepackage{url}            % simple URL typesetting
\usepackage{booktabs}       % professional-quality tables
\usepackage{amsfonts}       % blackboard math symbols
\usepackage{nicefrac}       % compact symbols for 1/2, etc.
\usepackage{microtype}      % microtypography
\usepackage{xcolor}         % colors
\usepackage{graphicx}

\title{R\&B - Rhythm and Brain: Cross-subject Decoding of Music from Human Brain Activity}

\author{%
  Matteo Ferrante\thanks{These authors contributed equally to this work} \\
  Department of Biomedicine and Prevention\\
  University of Rome Tor Vergata\\
  \texttt{matteo.ferrante@uniroma2.it} \\
  \And
  Matteo Ciferri* \\
  Department of Biomedicine and Prevention\\
  University of Rome Tor Vergata\\
  \texttt{matteo.ciferri@students.uniroma2.eu} \\
  \AND
  Nicola Toschi \\
  Department of Biomedicine and Prevention\\
  University of Rome Tor Vergata\\
  A.A. Martinos Center for Biomedical Imaging\\
  Harvard Medical School/MGH, Boston (US)\\
}

\begin{document}

\maketitle

\begin{abstract}
Music is a universal phenomenon that profoundly influences human experiences across cultures. This study investigates whether music can be decoded from human brain activity measured with functional MRI (fMRI) during its perception. Leveraging recent advancements in extensive datasets and pre-trained computational models, we construct mappings between neural data and latent representations of musical stimuli. Our approach integrates functional and anatomical alignment techniques to facilitate cross-subject decoding, addressing the challenges posed by the low temporal resolution and signal-to-noise ratio (SNR) in fMRI data. Starting from the GTZan fMRI dataset, where five participants listened to 540 musical stimuli from 10 different genres while their brain activity was recorded, we used the CLAP (Contrastive Language-Audio Pretraining) model to extract latent representations of the musical stimuli and developed voxel-wise encoding models to identify brain regions responsive to these stimuli. By applying a threshold to the association between predicted and actual brain activity, we identified specific regions of interest (ROIs) which can be interpreted as key players in music processing. Our decoding pipeline, primarily retrieval-based, employs a linear map to project brain activity to the corresponding CLAP features. This enables us to predict and retrieve the musical stimuli most similar to those that originated the fMRI data. Our results demonstrate state-of-the-art identification accuracy, with our methods significantly outperforming existing approaches. %Our findings highlight the potential for neural-based music retrieval systems, opening new avenues for personalized music recommendations and therapeutic applications. Future work could explore higher temporal resolution neuroimaging methods and generative models to further enhance decoding accuracy and explore the neural underpinnings of music perception and emotion. 
Our findings suggest that neural-based music retrieval systems could enable personalized recommendations and therapeutic applications. Future work could use higher temporal resolution neuroimaging and generative models to improve decoding accuracy and explore the neural underpinnings of music perception and emotion.

\end{abstract}

\section{Introduction}

Music universally permeates cultures, exerting a profound influence on the lives of those who perceive its harmonies and rhythms. Despite its pervasive role, the intricacies of how music impacts the human brain remain enigmatic. Music engages complex neurological pathways, triggering diverse emotional responses, evoking vivid episodic memories, and even interacting with various neurological disorders. These interactions suggest a deep and multifaceted relationship between music and brain function, warranting extensive scientific exploration \citep{Margulis2019}. This study investigates the extent to which music can be decoded from human brain activity measured with functional MRI (fMRI).

Historically, the study of how the brain interprets and processes music has been a topic of classical inquiry within neuroscience \citep{raglio_music_2019}. However, recent advancements have revolutionized this field, making it practicable to use AI to explore and decode brain patterns relative to a wide set of stimuli \citep{oota2023deep}. In this context, the emergence of extensive datasets coupled with robust, pre-trained computational models presents an unprecedented opportunity. These tools enable us to construct detailed mappings between neural data and the latent, compact representations of external stimuli, such as images \citep{ferrante2023semantic, ferrante2023brain, ozcelik2023braindiffuser, chen2022seeing, scotti2023reconstructing}, videos \citep{chen2023cinematic}, language \citep{antonello2023scaling, Defossez2023DecodingSpeech, tang_semantic_2023_semantic_language}, and notably, music \citep{denk2023brain2music}. These works propose several retrievals as well as generative pipelines to create a map between neural data and latent representations of external stimuli. The neural data is primarily measured via functional magnetic resonance imaging (fMRI), magnetoencephalography (MEG), or electroencephalography (EEG), and the latent representations are commonly obtained from large pretrained models. The estimated latent representations are further used for stimulus retrieval or conditioning of a generative model to generate e.g. images in vision decoding. Typically, these pipelines involve linear mappings between these two spaces (brain and latent representations of stimuli) and require subject-specific models, although some approaches to multisubject brain representations or alignment and nonlinear mappings exist \citep{ferrante2023eyes, benchetrit2023brain, scotti2024mindeye2}.

\begin{figure}
\centering
\includegraphics[width=\linewidth]{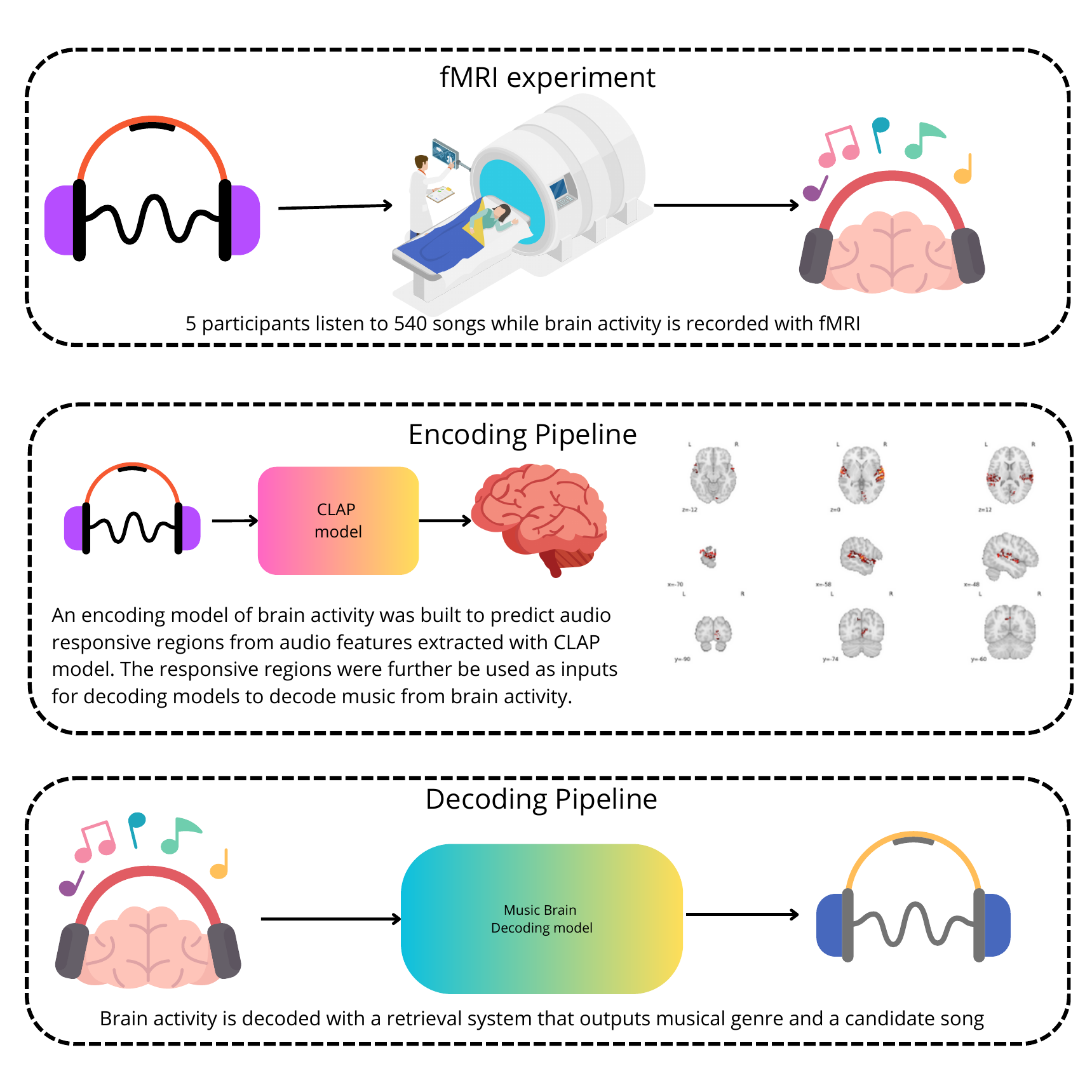}
\caption{Overview of our pipeline. \textbf{Top pane}: In the GTZan fMRI experiment, five participants were exposed to auditory stimuli that included multiple musical tracks while their brain activity was monitored via functional MRI. This setup captures the direct neural response to complex auditory inputs. In the \textbf{middle pane}, our encoding pipeline is described: Starting from the music stimulus, we first obtain its latent representation using the CLAP model. Subsequently, we develop voxel-wise encoding models to map the brain's response to these stimuli to this latent space. A threshold is then applied to the voxel-wise correlation between real and predicted brain activities to identify brain regions whose activity allows the best decoding of musical stimuli. These regions are considered as most responsive to music-related regions of interest (ROIs). The \textbf{bottom pane} outlines our decoding pipeline, which is primarily retrieval-based. We train a model that inputs brain activity from the previously identified ROIs and predicts the corresponding CLAP features. Using these features, we then search within the CLAP latent space for the closest musical stimulus, selecting the nearest k (k=5) stimulus as our retrieved samples.}

\label{fig:scheme}
\end{figure}

Understanding these complex relationships is both fascinating and informative, potentially offering insights into fundamental brain functions. For example, understanding the connection between music perception and neural responses could unlock novel avenues for diagnosing and treating neurological disorders. Moreover, it could enhance music therapy approaches, potentially leading to innovative treatments that harness the therapeutic properties of music \citep{Kamioka2014-js, De_Witte2022-kh}.

In this work, we aim to decode music from brain activity—a process that involves translating the neural signals evoked by music into a comprehensible format. This objective challenges us to retrieve complex auditory information encoded within the brain’s activity. In the case of fMRI, the primary challenge lies in decoding a signal of inherently higher frequency than the neural signal, \textcolor{black}{which is further confounded by the local variation in the brain of the Haemodynamic Response Function (HRF)}. Additional limitations include the constraints posed by small datasets typically comprising few subjects with intrinsic between-subject anatomical and functional differences.

To address these challenges, we first constructed encoding models to identify brain regions responsive to musical stimuli. We then aggregated brain activity across subjects to facilitate a cross-subject decoding approach. This included aligning functional brain data and mapping the identified regions’ activity to the latent representations of music stimuli. These representations were derived using an open-source, multimodal pre-trained foundation model known as Contrastive Language-Audio Pretraining (CLAP) \citep{elizalde2022clap}. In the final stages of our study, we compared the representations of music estimated from brain data with their true counterparts, employing a selection criterion that identified the five closest matching representations as potential candidates for accurate decoding.

The studies most closely related to our research include \cite{Bellier2023} and \cite{denk2023brain2music}. \cite{Bellier2023} demonstrate that time-frequency decompositions can be effective representations for this type of task, and that they can be performed using both linear and nonlinear approaches to decode the auditory experience using invasive iEEG data. 

Another pivotal study, \citep{denk2023brain2music}, shares similarities with our approach in that it addresses the challenges of retrieval-based as well as generative music decoding using the same fMRI dataset we employ here. However, unlike our methodology, \cite{denk2023brain2music} uses subject-specific decoding pipelines based on anatomical atlases and proprietary models like MuLAN and MusicLM \citep{agostinelli2023musiclm, huang2022mulan}.

In this paper, we advance the state of the art by designing a streamlined pipeline that leverages open-source models. Our approach begins by identifying brain regions whose activity can be reliably modelled using latent representations of audio stimuli. Subsequently, we use the brain activity from these regions to construct cross-subject decoding pipelines. Figure~\ref{fig:scheme} depicts our pipeline. In our work, we aspire to refine our understanding of how music is processed within the brain and to lay the groundwork for future explorations into the therapeutic potential of music in neurological settings.

\section{Material and Methods}
In this section, we describe the proposed method and the data we used. The data are publicly available and can be requested at \url{https://openneuro.org/datasets/ds003720/versions/1.0.1}. All experiments and models were trained on a server equipped with four NVIDIA A100 GPU cards (80GB RAM each connected through NVLINK) and 2 TB of System RAM. Throughout this paper, we will use the terms "fMRI data" as "brain activity", "neural activity" or "neural representations" interchangeably. These terms all stand for the fMRI signal, averaged over the time-points related to a specific stimulus, i.e. a 3D map. Additionally, the terms "musical features" or "musical representations" always refer to the embedding of musical stimuli generated by the CLAP model.
Code is available at this repository: \url{https://github.com/neoayanami/fmri-music-retrieve}.

\subsection{Data}
The GTZan fMRI dataset \citep{ds003720:1.0.1} comprises functional magnetic resonance imaging (fMRI) data collected from five subjects ("sub-001" to "sub-005") while they listened to music stimuli drawn from 10 distinct genres. The experimental protocol included 18 fMRI acquisitions (i.e. "runs") per subject, consisting of 12 training runs and 6 test runs. Each run is also associated with detailed information about each stimulus, including onset time, genre type, track name, and start and end times of excerpts from the original music stimuli. All stimuli have a duration of 15 seconds, including 2 seconds of fade-in and fade-out (a total of 4 seconds). The data are provided in intensity normalized form, i.e. after root mean square (RMS) normalization. In the test run ensemble, each musical stimulus was administered four times and the brain activity averaged across identical stimuli. Data averaging improves the signal-to-noise ratio (SNR) and enhances the detection of consistent neural responses associated with the stimulus under investigation. 

After motion correction, we co-registered the fMRI data to the Montreal Neurological Institute (MNI) standard space using a T1w anatomical image as reference for each subject, and applied detrending and standardization at the run level.  The final step involved "delaying" the brain activity by 3 Repetition Times (TR) (i.e. 4.5 s) in order to account for the peak of the hemodynamic response, and averaging the following 15 seconds to obtain a neural representation for each musical stimulus. Our final dataset is therefore composed of a total of 540 stimuli-processed fMRI pairs for each subject, divided into 480/60 train/test, as defined by the authors of the dataset.  We used FSL \citep{Jenkinson2012} for co-registration and the Nilearn python library \citep{nilearn} to perform all other preprocessing steps.

\subsection{Functional Alignment}
To address the inherent variability in brain structure/function across different individuals, we explored three distinct methodologies for aggregating cross-subject data. These techniques aim to enhance the robustness and accuracy of decoding models by aligning and integrating neural data from multiple subjects. Each method offers a unique approach to the challenge of intersubject variability, a common hurdle in neuroimaging studies.

The first method we implemented was anatomical alignment, which uses standard brain atlases to align brain imaging data from different subjects based on their anatomical landmarks. By mapping each subject's data to a common anatomical space, we can directly compare and combine data across individuals, despite differences in brain size, shape, or orientation. This method is widely used in neuroimaging as it facilitates the direct comparison of localized brain activity across subjects. 

Moving beyond mere anatomical correspondence, our second method, functional alignment, aligns brain activity based on functional data. This technique involves matching brain regions that exhibit similar activity patterns during specific tasks or stimuli across different subjects. Unlike anatomical alignment, functional alignment accounts for individual variations in brain function topology that may not align with variations in physical brain structures, making it particularly advantageous for studies where functional responses to complex stimuli are the primary focus. To this end, we leveraged the "hyperalignment" strategy proposed by \cite{Haxby2011-up} based on Procrustes analysis.

Lastly, given recent literature \citep{ferrante2023eyes, defossez2023decoding, benchetrit2023brain} which demonstrated that linear layers are a useful tool to align neural representations into a common space, we employed ridge regression to aggregate cross-subject brain data. This approach applies regularization to address multicollinearity in high-dimensional datasets, which is typical of fMRI data. By introducing a penalty term, ridge regression combines voxel-wise data from different subjects into a unified model while enhancing the stability and generalizability of our predictions. Each of these methods was tested for its potential to improve the accuracy of our decoding models, with the goal of establishing a reliable approach to interpreting complex brain data in a multi-subject context.

\subsection{Music Feature Extraction}
Our brain engages with music in intricate, non-linear ways, forming representations that support our cognitive processes. This complexity suggests that a multimodal pre-trained model like CLAP, \cite{elizalde2022clap}) may mimic some aspects of how our brains process music.  Under this hypothesis, CLAP can transform musical stimuli into a vectorial representation that could present topological similarities similarity with the brain representations, allowing the identifications of simple mapping between the latent representations generated by CLAP and those generated by the human brain.

CLAP is a multimodal neural network designed for contrastive learning in the realm of audio and text processing. It is trained on a diverse set of audio and text pairs, learning to align text and audio latent representations. The model employs the SWINTransformer \citep{liu2021swin} to extract audio features from log-Mel representations and the RoBERTa model \citep{liu2019roberta} to extract text representations, both projected into a shared latent space of identical dimensionality. The similarity between audio and text features is measured using cosine similarity.

\begin{figure}
  \centering
  \includegraphics[width=\linewidth]{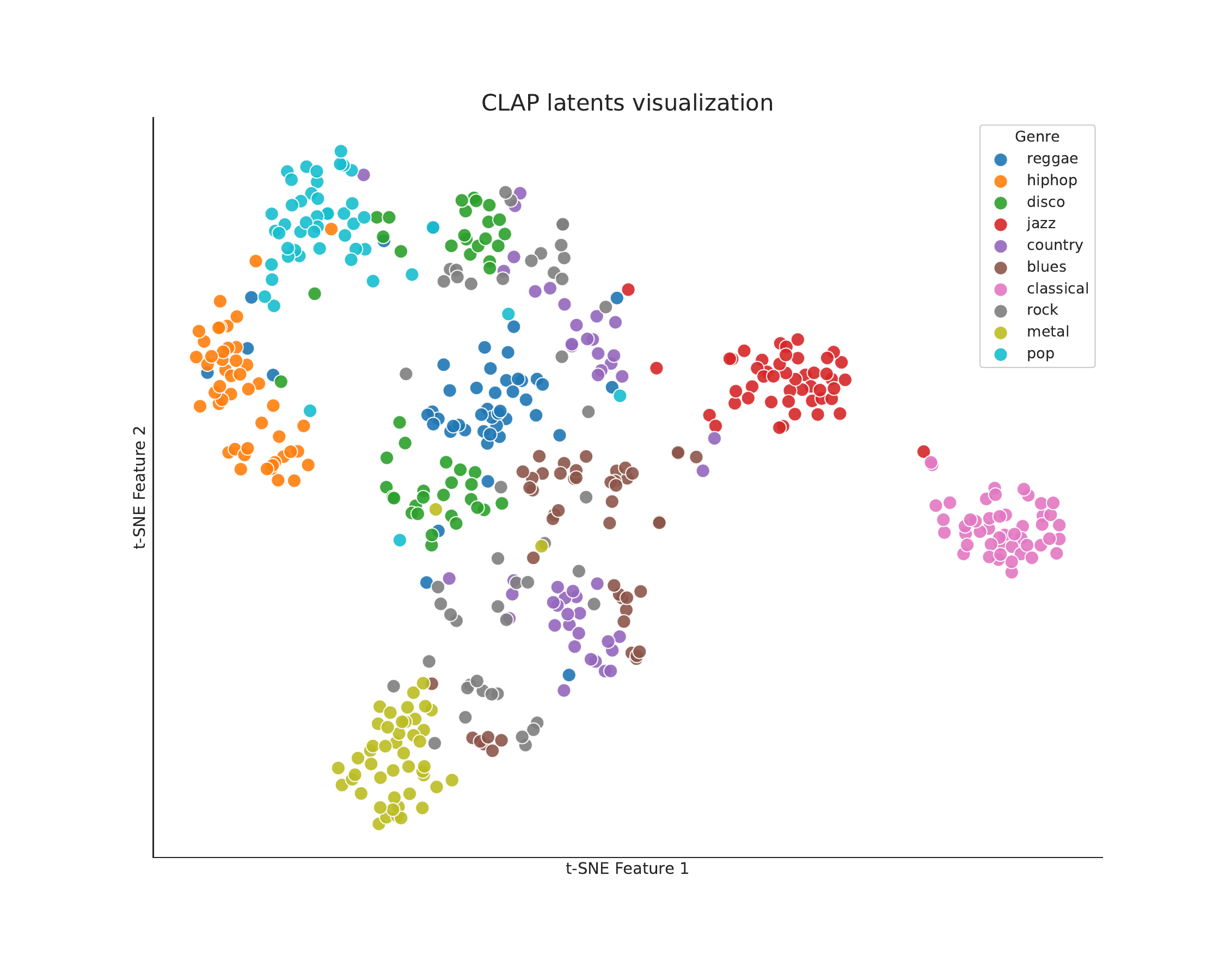}
  \caption{Two-dimensional t-SNE representation of CLAP latent representations of music, coloured by different musical genres.}
  \label{fig:tSNE}
\end{figure}

Figure \ref{fig:tSNE} shows the results of using t-Distributed Stochastic Neighbor Embedding (t-SNE, \cite{tsnevandermaaten08a}) to create a 2D visualization of the true music features overlayed on genre labels, offering a qualitative understanding of how the CLAP model's representations are able to separate different genres.

\subsection{Encoding Models}
The primary goal of this part of our study was to identify brain regions responsive to musical stimuli by constructing voxel-wise encoding models. These models map the latent representations of musical stimuli onto voxel-wise brain activity. To assess the efficacy of each voxel's model, we employed a cross-validation scheme, wherein the correlation between the predicted and real brain activities of each voxel was measured.

Model training incorporated a hyperparameter search for the regularization parameter $\alpha$. We explored a range of $\alpha$ values set on a logarithmic scale from $10^{-2}$ to $10^{3}$. Upon completing the model training, we established an empirical threshold for selection at a correlation of 0.1. This threshold was empirically chosen during preliminary explorations and was used to generate a mask of the brain regions. This mask delineates areas showing higher responsiveness to musical stimuli.

\subsection{Decoding Model}
Following the identification of brain regions responsive to music, our next objective was to construct a common model that could map the brain activity from these regions to the latent representations of musical features.  This model aims to facilitate a translation process where the neural responses could potentially be directly mapped into musical features, by creating a predictive model where the brain's response could serve as a proxy for the music itself, also illustrating a direct link between neural activity and musical perception. To this end, we trained a Ridge regression with hyperparameter optimization between the aligned brain activity of all subjects in "music-responsive" brain regions. Successively, we then focused on optimizing the retrieval process within the testing dataset. For each predicted musical feature, we selected the top-k closest elements based on the lowest L2 (Euclidean) distance between predicted and true musical features in CLAP space. This approach forms the basis of a straightforward retrieval pipeline, where the model searches for and retrieves the most similar musical stimuli from the latent space, based on the neural activity they elicited.

\subsection{Evaluation}
In our study, we measured the identification accuracy as described in the Brain2Music framework \citep{denk2023brain2music}. Identification accuracy quantifies how accurately the predicted \(d\)-dimensional features correspond to the target features by computing the Pearson correlation coefficient between each pair of predicted and target features. In our case, the features are the estimated and true CLAP features (last layer, dimensionality 512). The accuracy for each prediction is the proportion of correct identifications, where a correct identification occurs if the correlation (computed as above) for a given prediction is higher than the one for any other prediction. In detail, the metric is calculated as follows: first, construct a correlation matrix between the predicted and true embeddings.  Each element of this matrix, \( C_{i,j} \), represents the Pearson correlation coefficient between the \( i \)-th predicted embedding and the \( j \)-th target embedding. For each predicted embedding, determine whether the correlation with its corresponding target (diagonal element \( C_{i,i} \)) is greater than the correlations with all other targets (non-diagonal elements \( C_{i,j} \) for \( j \neq i \)). The identification accuracy for each prediction is then calculated using an indicator function:
\[
  \text{id\_acc}_i = \frac{1}{n-1} \sum_{j=1}^n 1 \left[ C_{i,i} > C_{i,j} \right]
\]
where \( 1[\cdot] \) is the indicator function that returns 1 if the condition is true and 0 otherwise. The formula ensures that each comparison excludes the self-comparison (\( j = i \)). The overall identification accuracy is the average across all predictions:
\[
  \text{id\_acc} = \frac{1}{n} \sum_{i=1}^n \text{id\_acc}_i
\]

Identification accuracy is especially useful in scenarios where the data may lead to ambiguous interpretations, requiring robust model performance to correctly identify the underlying condition or stimulus. Following an intuitive explanation of identification accuracy provided in \citep{denk2023brain2music} adapted for our case: from a practical perspective, consider a model that achieves an identification accuracy of 90\%. This implies that, on average, 10\% of the predictions are incorrect, i.e. cases where another candidate (not the correct "target") corresponds to a higher correlation coefficient than the correct candidate. In a dataset containing 60 examples, this would mean that the correct music track, on average, is ranked sixth (10\% of 60 equals 6) in terms of correlation, suggesting that five other music stimuli were mistakenly rated as more likely candidates as compared to the correct one.

For demonstration purposes, we provide qualitative examples of decoded music. These examples can be accessed at the provided URL \url{https://mind2music.my.canva.site/decoding-music-from-brain-activity-exploring-the-neural-correlates-of-music-perception}, where listeners can directly experience the output of our decoding process, offering an auditory validation of the model's performance.

\section{Results}
This study examined the effectiveness of various embedding models and functional alignment strategies in identifying and classifying musical genres based on brain activity data. The results highlight significant advancements in genre classification accuracy and provide insights into the spatial distribution of musically responsive brain regions.

\subsection{Encoding Models and Delineation of brain areas responsive to music}
\begin{figure}
  \centering
  \includegraphics[width=\linewidth]{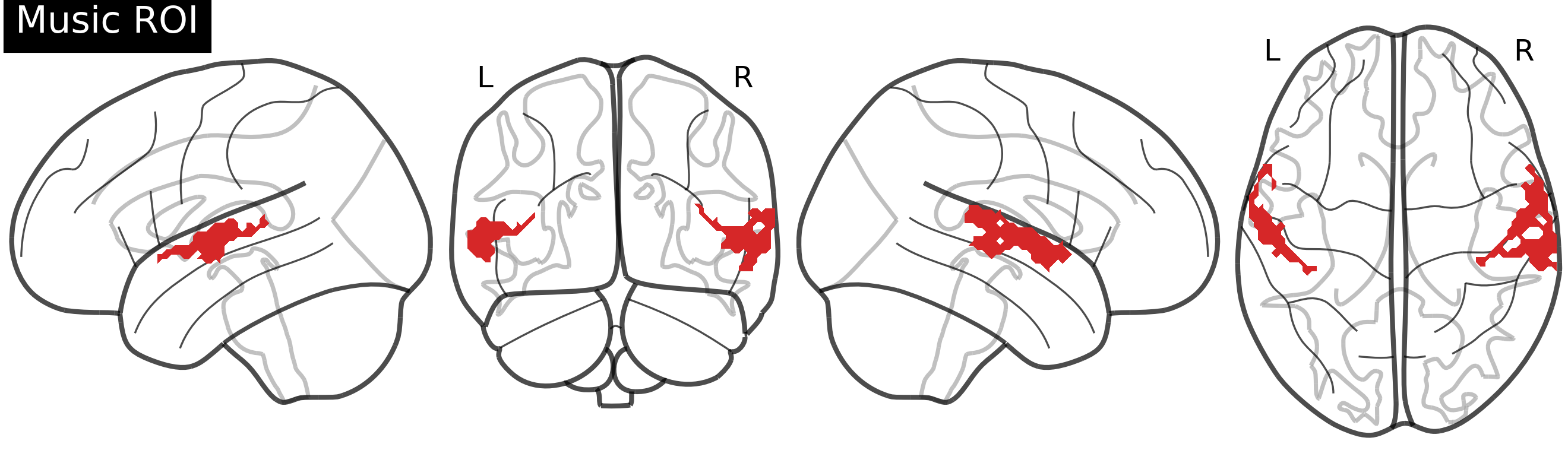}
  \caption{Regions of interest (ROIs) corresponding to musically responsive areas were identified by applying a threshold to the correlations between predicted and actual brain activity. This process was part of a cross-validation procedure used in the encoding models.}
  \label{fig:ROI}
\end{figure}

By setting a threshold of 0.1 (see methods), the encoding models identified 833 voxels in total. This threshold was empirically determined to optimize the balance between sensitivity and specificity in our voxel selection procedure. Figure \ref{fig:ROI} shows the distribution of the relevant voxels within anatomical brain space, which appear to co-localize within lateral and temporal regions.

\subsection{Identification Accuracy}
As shown in Table \ref{tab:test_identification_accuracy}, our proposed methods with functional alignment techniques, denoted \textit{linear} and \textit{hyperalign}, demonstrated superior performance with identification accuracies of 0.9012 ± 0.01573 and 0.8805 ± 0.0231, respectively, outperforming other baselines and the anatomical alignment method. The linear alignment method, in particular, shows the highest performance, underscoring the efficacy of our linear modelling approach to achieve cross-subject music decoding from brain activity. This is in accordance with our previous observation in vision decoding \citep{ferrante2023eyes}.

\begin{table}[ht]
\centering
\caption{Comparison of Test Identification Accuracy}
\label{tab:test_identification_accuracy}
\begin{tabular*}{\textwidth}{@{\extracolsep{\fill}}lc@{}}
\toprule
\textbf{Embedding} & \textbf{Test Identification Accuracy} \\
\midrule
SoundStream-avg & $0.674 \pm 0.016$ \\
w2v-BERT-avg & $0.837 \pm 0.005$ \\
MuLan\textsubscript{text} & $0.817 \pm 0.014$ \\
MuLan\textsubscript{music} & $0.876 \pm 0.015$ \\
\midrule
Ours - anatomical & $0.7746 \pm 0.01551$ \\
\textbf{Ours - hyperalign} & $\mathbf{0.8805 \pm 0.0231}$ \\
\textbf{Ours - linear} & $\mathbf{0.9012 \pm 0.01573}$ \\
\bottomrule
\end{tabular*}
\end{table}

\subsection{Genre Decoding}
\begin{figure}
  \centering
  \includegraphics[width=.80\linewidth]{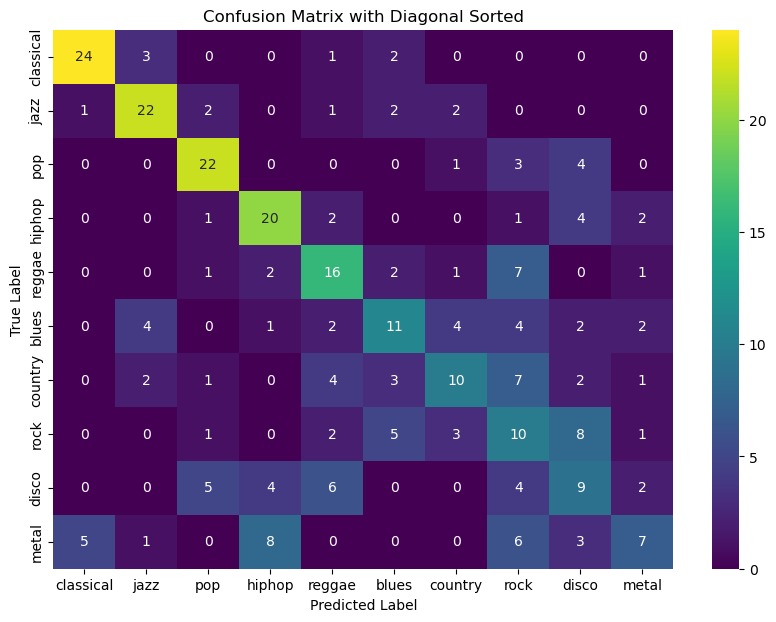}
  \caption{Confusion matrix showing our model's accuracy (number of correct predictions over the number of total predictions) in classifying musical genres based on fMRI data from five participants. Diagonal elements represent correct predictions for each genre, while off-diagonal elements indicate misclassifications. Each genre has 30 music stimuli, evenly distributed across the subjects; a value of 30 in the main diagonal therefore represents 100\% accuracy. The model performs well for classical, jazz, and pop genres, with minimal confusion, while disco and metal genres show higher misclassification rates, likely due to overlapping music features. The matrix highlights the effectiveness of the cross-subject decoding pipeline and areas for improvement.}
  \label{fig:confusion_matrix}
\end{figure}

\begin{figure}
  \centering
  \includegraphics[width=\linewidth]{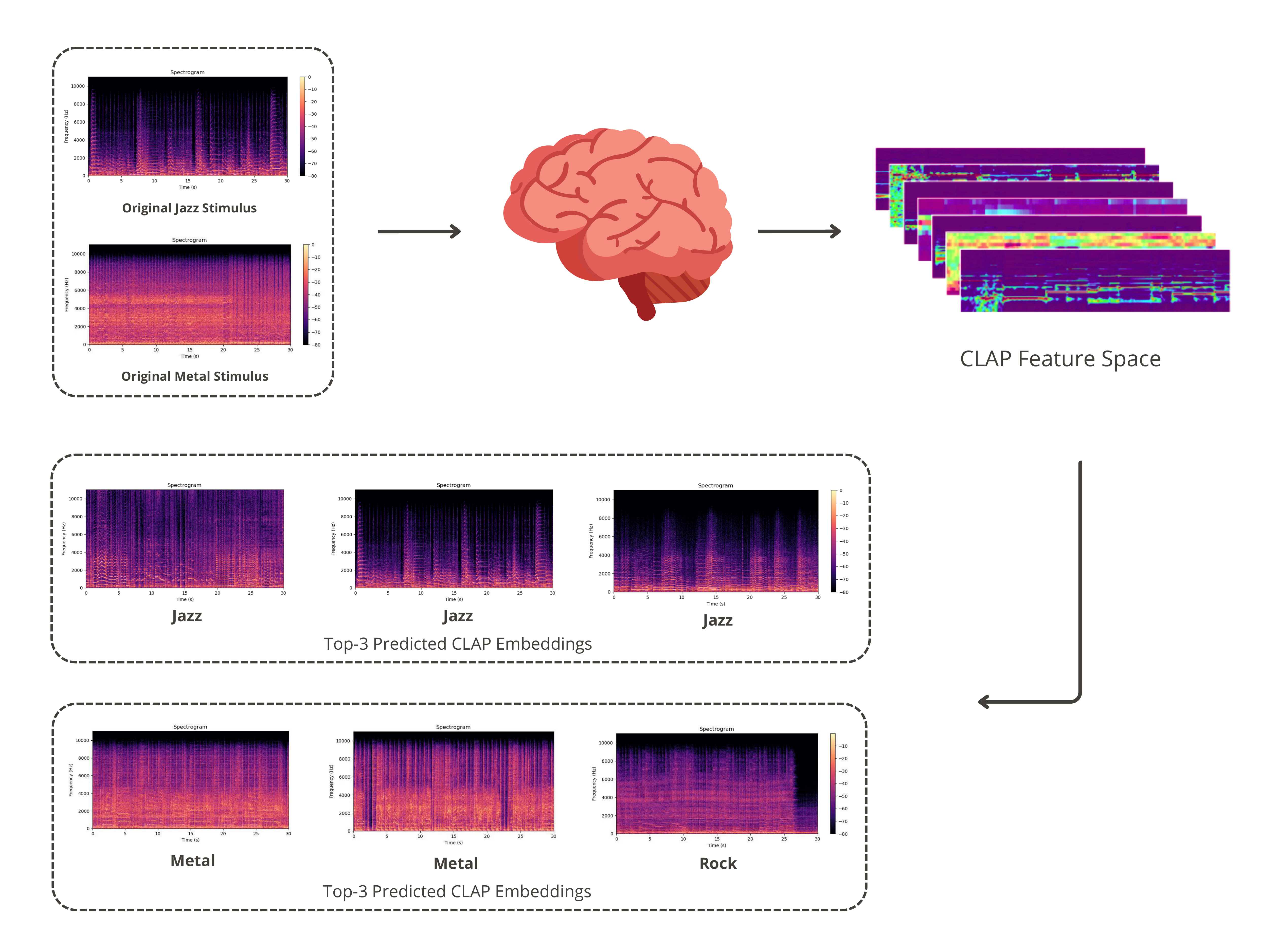}
  \caption{Time-frequency Decompositions (TFDs - used as illustrative visual aids to estimate similarity between audio data) of original musical stimuli (jazz and metal) and the stimuli decoded from the top-3 CLAP embeddings predicted using the Ridge regression decoding model. The left side displays the TFD  of the original jazz stimulus, while the right side shows the TFDs of the original metal stimulus. Below each original stimulus, the top-3 predicted stimuli are shown. For the jazz stimulus, the predicted simuli were all identified as jazz. For the metal stimulus, the top-3 predictions included two metal and one rock embedding. This comparison highlights the model's ability to accurately predict musical genres from brain activity, while also illustrating occasional genre misclassification, particularly in more complex or overlapping genre spaces.}
  \label{fig:spec_retr}
\end{figure}

The confusion matrix shown in Figure \ref{fig:confusion_matrix} illustrates the model's capability to classify musical genres based on brain activity, with a notable concentration of correct predictions along the diagonal. Classical and jazz genres showed high accuracy with minimal confusion, suggesting that they correspond to distinct neural representations. However, genres like metal and disco exhibited more confusion, potentially indicating less separability in the CLAP space. For example, the confusion between disco and metal may arise from similar rhythmic patterns or instrumentation that blur genre-specific boundaries in neural encoding. Figure \ref{fig:spec_retr} shows the similarity between the retrieved music and the original genre stimulus, using time-frequency as visual aids. Within the retrieved cluster, the exact stimulus is found very often, emphasizing the effectiveness of the pipeline. Given feature overlap, it is common to encounter different genres in the retrieved group of music stimuli compared to the stimulus, although always within genres that exhibit shared acoustic patterns.

\subsection{Impact of Functional Alignment techniques}
The choice of functional alignment techniques significantly enhanced the identification accuracy compared to baselines that did not make use of alignment. This improvement indicates that aligning functional brain data across subjects, while preserving individual differences in brain anatomy, allows for more accurate generalizations when decoding music genres from brain activity when compared to single-subject modelling. The technique effectively harnesses shared information across different subjects, thereby boosting the overall model's performance \citep{denk2023brain2music}.

Compared to existing studies, such as those using basic MuLan or SoundStream embeddings \citep{huang2022mulan,denk2023brain2music}, our method provides high performances in music track retrieval and genre classification accuracy. Previous studies often did not account for individual variations in brain anatomy and function as effectively, which our hyperalignment and linear methods address directly.

The results from this study not only reinforce the utility of advanced machine learning techniques in neuroscience but also pave the way for more personalized and accurate interpretations of brain activity in response to complex stimuli like music. Future work could explore deeper neural network architectures or alternative machine learning models that might further refine the accuracy of musical genre classification from brain imaging data.

\subsection{Decoding in Time}
\begin{figure}
  \centering
  \includegraphics[width=.77\linewidth]{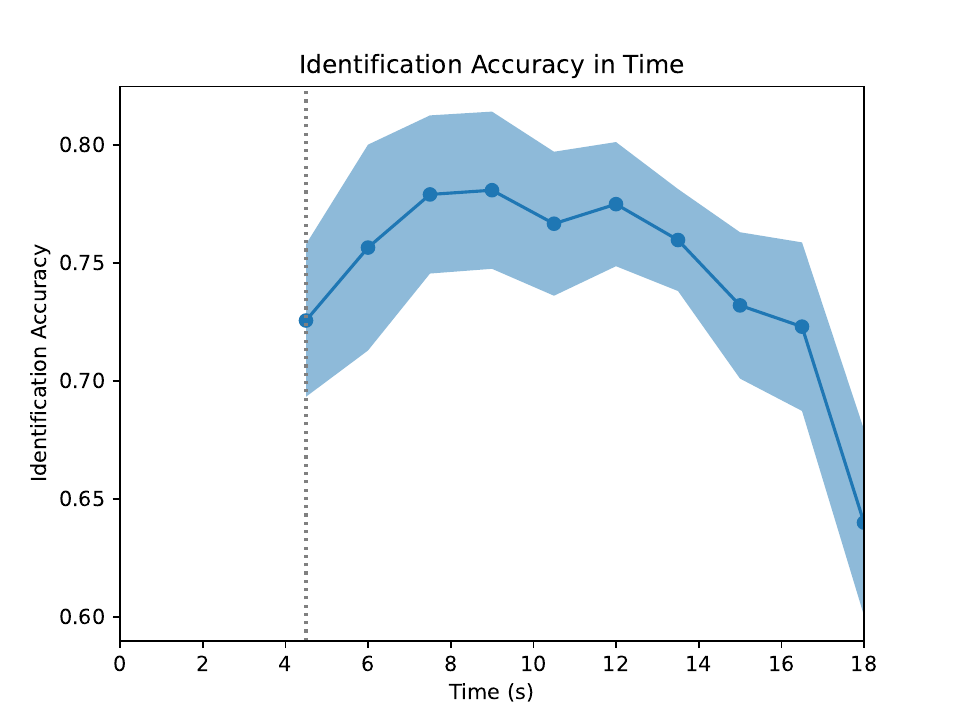}

\caption{Identification accuracy of the music decoding model over a time course of 18 seconds, skipping the first 4.5 s to account for HRF delay. The y-axis represents the identification accuracy, while the x-axis represents the time in seconds, where 0s indicates stimulus onset. There is an evident trend of increasing identification accuracy as time progresses, reaching a peak towards the later part of the time window. This indicates that the model's ability to accurately decode musical genres from brain activity improves with longer exposure to the musical stimuli, suggesting that prolonged neural engagement with the music enhances the decoding performance.}
  \label{fig:idacc_intime}
\end{figure}

In our main experiment, we averaged the 15 seconds of fMRI data for each musical stimulus. Another possible interesting research question is when, after the stimulus onset, a peak in performance for music decoding can be observed. To address this question, we evaluated the neural responses contained in each fMRI volume. This analysis relies on identical procedures as described above; however, instead of using averaged brain activity over 15s as input for the decoding model, instantaneous (i.s. sample-wise) brain activity is used, resulting in a decoding-in-time representation (Figure \ref{fig:idacc_intime}). By identifying the samples/time delays at which the identification accuracy is highest, this approach illustrates the specific temporal dynamics underlying music perception within the brain.

\section{Discussion}
The findings of this study provide compelling evidence that decoding music from cross-subject neural activity is not only feasible, but also remarkably accurate when appropriate computational approaches and neural data alignment techniques are employed and adapted. This opens up numerous possibilities for understanding the cognitive processing of music and its applications, ranging from therapeutic practices to advanced brain-computer interfaces.

\subsection{Implications of Music Decoding}

The successful decoding of music genres from brain activity suggests profound implications for cognitive neuroscience and psychological studies. By associating specific genres with distinct patterns of brain activation, researchers can further explore how these patterns correlate with cognitive functions, emotional states, and individual preferences. This understanding could eventually lead to personalized music interventions designed to manage various psychological conditions such as anxiety, depression, and stress. Further refinement of this process could lead to neural-guided recommendation systems, allowing individuals to receive personalized music suggestions based on neural similarities with music stimuli they enjoy or those that evoke specific emotions.

\subsection{Performance on Genre Decoding}
Our analysis achieved results in line with \citep{NAKAI2022107675},  further showing that certain genres like classical and jazz are more distinctly encoded in the brain, possibly due to their unique structural and rhythmic complexities which might engage specific neural pathways. However, the confusion between closely related genres like rock and metal highlights the challenges of distinguishing between potentially similar auditory stimuli and suggests a need for more refined modelling techniques that can capture subtle nuances in music perception.

\subsection{Identification of music-related brain regions}
Our results identified key brain regions involved in music perception and processing. Specifically, we identified the superior temporal gyrus (STG) \citep{Amusia}, primary auditory cortex \citep{Warren2008-pd}, planum temporale \citep{Warren2003-yq}, and potentially the inferior parietal lobule \citep{Amusia}. These areas are essential for decoding various aspects of auditory and musical stimuli, contributing to our ability to perceive and appreciate music. The superior temporal gyrus (STG), which includes the primary auditory cortex, is crucial for processing auditory information such as pitch, rhythm, and timbre. The primary auditory cortex, located within the STG, plays a fundamental role in detecting and discriminating sound frequencies, allowing us to discern different notes and rhythms in music \cite{Warren2008-pd}. This region's function is vital for understanding melodies and the basic structural components of music. Adjacent to the primary auditory cortex is the planum temporale, a region involved in higher-order auditory processing \cite{Warren2003-yq}. The planum temporale is asymmetrically larger in the left hemisphere, a feature associated with language dominance, but it also plays a significant role in music processing \cite{Warren2003-yq}. This area is crucial for discerning complex auditory patterns and structures, such as harmonies and musical sequences. The ability of the planum temporale to process these intricate auditory stimuli contributes to our cognitive understanding of music and its structural components. In addition to the STG and planum temporale, the inferior parietal lobule is implicated in the integration of sensory information from various modalities \citep{Pando-Naude2021}. This region contributes to spatial awareness of sounds, which is important for perceiving the spatial dynamics of music, such as the localization of instruments within a stereo field. The inferior parietal lobule also plays a role in attention and the processing of rhythmic elements, enhancing our ability to perceive musical tempo and timing \cite{Pando-Naude2021}. This integrative function is essential for experiencing music as a coherent and dynamic auditory event. Together, these regions form a network that facilitates different aspects of music perception. The superior temporal gyrus and primary auditory cortex are central to decoding the basic auditory properties of music \cite{Amusia, Warren2008-pd}, while the planum temporale supports higher-order processing and pattern recognition. The inferior parietal lobule's involvement in sensory integration and attention further enriches our ability to experience and appreciate the spatial and temporal dimensions of music. These interconnected brain regions work in concert to provide a comprehensive and nuanced understanding of music, enabling listeners to engage with its emotional and aesthetic qualities fully.

\subsection{Impact on Musical Therapy}
There are potential applications of this research in the field of musical therapy that could be significant. Making a step towards a better understanding of the neural underpinnings of how music influences emotion and cognition can aid in developing more effective therapeutic protocols. As highlighted in \citep{raglio_effects_2016, raglio_music_2019}, music therapy has been shown to have beneficial effects on various patient outcomes. While still in early stages, genre-specific neural decoding could tailor these therapies to individual needs, enhancing their effectiveness.  Music therapy has been utilized in various clinical settings, demonstrating positive outcomes in patients with conditions such as Alzheimer's disease, stroke, and depression \citep{Kamioka2014-js, De_Witte2022-kh}. By decoding how different genres affect brain activity, therapists could potentially customize music interventions that align more closely with the neural and emotional states of individual patients. This personalized approach could maximize therapeutic benefits by targeting specific neural circuits involved in emotional regulation and cognitive function. Moreover, further research into the relationship between music and neural responses could contribute to the development of innovative treatment modalities. For instance, integrating neurofeedback mechanisms that respond to real-time neural data could enable dynamic adjustments in musical stimuli, optimizing therapeutic outcomes. This approach could be particularly effective in managing chronic pain, stress, and anxiety, where music's role in altering brain states can be leveraged for long-term health benefits \citep{Koelsch2011-oi, Koelsch2014, koelsch_investigating_2006}. Understanding the specific neural mechanisms involved in music perception and emotional processing also provides insights into broader applications in cognitive neuroscience. For example, exploring how music can enhance cognitive rehabilitation in post-stroke patients or improve social communication skills in individuals with autism spectrum disorder represents promising research avenues. The ability to decode and harness the power of music at a neural level opens up new possibilities for both clinical practice and scientific inquiry into the profound effects of music on the human brain \citep{nakai_correspondence_2021, NAKAI2022107675}.

\subsection{Deeper Investigation of  Music and Emotions}
Further research could benefit from exploring the intricate connections between music and emotions, a relationship well-documented in the studies by \citep{koelsch_investigating_2006, Koelsch2011-oi, Koelsch2014}. By decoding the emotional content of music from brain activity, researchers could gain insights into the emotional processing in the brain, providing a clearer picture of the emotional impacts of music at a neurological level. Envisioning a significant advancement for the future, we could consider this type of research as the foundation for a neural recommendation system. This system could potentially offer personalized music track suggestions based on our emotional and neural states or even suggest music stimuli that could guide us toward new emotional experiences.

\subsection{Extension to Generative Music}
Looking forward, the decoding techniques used in this study could be extended to generative music systems, potentially leading to innovative applications in creating music from brain activity, including musical imagery.

At the time of writing, the primary reason we are focusing on retrieval rather than generation is the low temporal resolution of fMRI acquisition. This limitation constrains the possibility of generating music online based on neural dynamics, which however might be achievable with other neural activity measures like iEEG or MEG.  A particularly intriguing prospect is to replace the retrieval module with a generative stage, especially by combining music decoding with imagery. Imagine an artist entering the scanner and envisioning a music track to be decoded through this process. The resulting piece could be seen as a collaborative creation between the artist’s imagination and artificial intelligence, potentially giving rise to a new art form where learned musical priors are transformed and used by neural decoding models to produce unique artistic expressions. Such systems would not only deepen our understanding of the creative processes that underpin music generation but also open the door to innovative forms of artistic expression that are directly influenced by neural dynamics.

\subsection{Limitations}
Despite these advancements, several limitations remain. The neural signals used in this study are inherently noisy and are only a subsampled representation of brain activity, which limits the detail and accuracy of the music that can be reconstructed. Rhythmic elements, particularly those at fine temporal resolutions, remain challenging to decode accurately due to the limitations in the temporal resolution of fMRI technology. Moreover, the extensive scanning time required for collecting sufficient data is a practical limitation that could restrict the use of these techniques in everyday applications.

\subsection{Future Work}
Future research could explore the use of alternative neuroimaging methods, such as electroencephalography (EEG) or intracranial EEG (iEEG), which offer higher temporal resolution and could potentially provide more detailed insights into the neural encoding of music. Additionally, the development of more sophisticated generative models that can better handle the complexity and variability of neural data represents a promising direction for both academic research and practical applications in neuromusicology.

\section{Conclusion}

This study demonstrates high identification accuracy in decoding music from cross-subject neural activity using a streamlined retrieval pipeline, setting a new benchmark in neuromusicology with significant implications for therapeutic and personalized music applications.

\bibliographystyle{plainnat}
%\bibliography{Styles/neurips_2024}
\bibliography{neurips_2024}

\end{document}